# Aggregation and settling in aqueous polydisperse alumina nanoparticle suspensions

Sanjeeva Witharana*, Chris Hodges, Dan Xu, Xiaojun Lai, Yulong Ding

*Institute of Particle Science and Engineering, University of Leeds, Leeds, LS2 9JT, United Kingdom*

*Corresponding author: Phone: +441133432543 Email: switharana@ieee.org

**Abstract**

Nanoparticle suspensions (also called nanofluids) are often polydisperse and tend to settle with time. Settling kinetics in these systems are known to be complex and hence challenging to understand. In this work, polydisperse spherical alumina ($Al_2O_3$) nanoparticles in the size range of ~10-100nm were dispersed in water and examined for aggregation and settling behaviour near its isoelectric point (IEP). A series of settling experiments were conducted and the results were analysed by photography and by Small Angle X-ray Scattering (SAXS). The settling curve obtained from standard bed height measurement experiments indicated two different types of behaviour, both of which were also seen in the SAXS data. But the SAXS data were remarkably able to pick out the rapid settling regime as a result of the high temporal resolution (10s) used. By monitoring the SAXS intensity, it was further possible to record the particle aggregation process for the first time. Optical microscopy images were produced on drying and dried droplets extracted from the suspension at various times. Dried deposits showed the rapid decrease in the number of very large particles with time which qualitatively validates the SAXS prediction, and therefore its suitability as a tool to study unstable polydisperse colloids.

# Introduction

Nanoparticles have gained wide recognition in variety of commercial and industrial applications over the years such as structural applications, skincare products, information and communication technology, biotechnology, and environmental monitoring instrumentation.

Alumina ($Al_2O_3$) in particular possesses a variety of commercial and industrial uses and has become one of the most important commercial ceramic materials (Uyeda 1991; Ichinose 1992). As a widely studied nanomaterial, $Al_2O_3$ nanoparticles have been applied in catalysis (Jodin et al. 2006; Cai et al. 2009; Yu et al. 2009), nanocomposites (Gleiter 2000; Bertsch et al. 2004), polymer modification (Cho et al. 2006), functionalization of textiles (Vigneshwaran et al. 2007), heat transfer fluids (Ding et al. 2010), and waste water treatment (Sharma et al. 2008). In addition, $Al_2O_3$ nanoparticles have featured in biological applications such as biosensors (Liu et al. 2011), biofiltration and drug delivery (Monteiro-Riviere et al. 2010), antigen delivery for immunization purposes (Skwarczynski and Toth 2011) and bactericides (Sadiq et al. 2009). Particle aggregation and settling are important in many of these applications and require further exploration.

Solid-liquid separation of particles has its own importance in a range of industrial and natural processes such as waste-water treatment, mineral separation, and deposition of sediments in rivers and lakes (Bergstrom 1992). Very often these processes involve suspensions that are flocculated and not colloidally stable. Moreover most industrial slurries contain a wide variety of particle sizes and shapes. For the case of a single particle in a static fluid (unhindered settling), the settling rate has been found to depend on the density and viscosity of the fluid as well as the density, size, shape, roundness, and surface texture of the particle (Dietrich 1982; Eckert et al. 1996). For the multi-particle case the situation is more complicated. Ayoub et al (1983) studied the settling of polydisperse spherical and non-spherical glass particles in oil with a particle loading of 20 – 50 wt%. The spherical particles were quicker to settle than non-spherical particles, and larger spheres were found to settle faster than smaller ones. At high concentrations (> 40 wt%), the polydisperse spheres (of diameter ~100μm) attained the theoretical settling rate expected from monodisperse spheres (unhindered settling). Jimenez and Madsen (2003) developed a formula to predict the settling rates for particle sizes between 63 μm and 1 mm in natural sediments. Knowledge of the particle shape is crucial in order to use this formula accurately. To simplify the use of the formula for non-spheres, a sediment shape factor and roundness were introduced. In a study of $TiO_2$ (17 – 41 vol%)-water slurries, Turian et al.

(1997) saw their settling data deviate significantly from the expected hindered settling correlations. The reason for the deviation was found to be the agglomeration of the 0.7µm $TiO_2$ particles during the settling process. The authors concluded that the existing correlations for hindered settling only accurately predicted the behaviour of monodisperse systems. In the absence of a suitable predictive correlation for all particle types, experimental methods have become the usual way to obtain the settling rates of complex polydisperse and aggregating systems.

The conventional method to measure settling rates is to fill a measuring cylinder with the slurry, mix it well to disperse the particles completely, and then measure the descent of the solid-liquid interface as a function of time (Turian et al. 1997; Zhu et al. 2000). However this method can introduce sizable errors in rapidly settling slurries since the settling interface is not easily identifiable and may change too rapidly for the naked eye to monitor. Other techniques previously used for settling rate measurement include laser scattering techniques, X-ray and γ-ray radiation, radioactive and magnetic tracers, ultrasonic velocimetry, magnetic resonance imaging (MRI), hydrometers and pressure sensors (Williams et al. 1990).

Pedocchi and Garcia (2006) discuss a laser scattering technique known as LISST (Laser In Situ Scattering Transmissometry) for use in settling experiments. LISST measures the suspension size distribution evolution over a period of time. The settling velocity is subsequently calculated from the evolution of each size class at the bottom of the settling column. Understandably this technique is restricted to suspensions that are not opaque, and is less accurate for very polydisperse suspensions.

Ultrasonic techniques such as UVP offer advantages at higher particle concentrations when the colloids are opaque. At very high volume fractions however this method becomes unreliable, since the solid phase increasingly hinders the propagation of the ultrasonic wave through the sample. Moreover this method can only track changes around the transducer plane, not the progressive sedimentation with time (Hunter et al. 2011). Magnetic Resonance Imaging (MRI) is non-invasive, works in opaque environments and with a wider range of solid concentrations. However, when 2-dimensional image measurements are required, a long acquisition time is needed for high resolution images. Therefore MRI is typically prescribed for systems with slow settling rates (Turney et al. 1995).

X-ray and γ-ray techniques are traditionally limited by the degree of attenuation measured as well as the acquisition time for each measurement. To date these techniques have been employed mainly for slow-settling slurries (Bergstrom 1992). Small Angle X-ray Scattering (SAXS) has previously been used to determine the size distribution and shape of nanoparticles and to study their aggregation under given conditions (Lee et al. 1998; Beaucage et al. 2004; Kammler et al. 2004; Chen et al. 2008; Seekkuarachchi and Kumazawa 2008a,b). These studies typically considered particles with a low degree of polydispersity and stable suspensions. Until recently the SAXS measurements were quite slow, requiring many minutes for each data point. However, the recently established synchrotron source at Diamond Light Source in UK is capable of producing a much higher number density of X-rays that allows the SAXS measurement time to be significantly reduced, in the order of 10 seconds per data point. This unprecedented speed enables the use of SAXS to examine rapidly settling colloids for the first time, and hence to fill the long-awaited void for an experimental technique to completely study such systems. Present work is aimed at fulfilling these aspirations.

This paper presents the findings from a series of experiments aimed at understanding the aggregation and settling behaviour of rapidly settling polydisperse alumina-water suspensions. Alumina particle concentration of 0.5 wt% was chosen for these experiments, as this is close to the upper limit for interparticle interference in light scattering techniques and hence demonstrates the usefulness of our approach. The settling rates are determined by both the conventional bed height method and the SAXS. Optical microscopy studies were carried out on evaporating droplets in an attempt to study the uniformity of the aggregates across the bed.

## Materials and Methods

### (a) Sample preparation

Nanotek dry alumina nanopowder was purchased from Nanophase Inc., IL, USA. According to the supplier the powder was produced by chemical vapour deposition and had a mean particle size of 46nm. Further details of the manufacturing process were classified as proprietary. Transmission electron microscopy (TEM) images taken by the authors on the as-received dry alumina powder (Figure 1) demonstrated wide polydispersity with particle sizes ranging from a few nanometres to > 100 nm. All of the individual nanoparticles were observed to be quite spherical in shape. Using this nanopowder aqueous 0.5 wt% alumina suspensions were prepared. Preparation of the suspension was as follows: A measured quantity of dry alumina nanopowder was gently added to deionised water (pH 7) while being magnetically stirred. After the solution cleared, ultrasound was applied for 4 hours. At intervals during this process samples were extracted from this suspension and measured by a Zeta-sizer Nano (Malvern

Instruments). There was approximately a 50% reduction in the measured mean particle size (Z-average). The as-prepared suspension was mildly acidic at pH 4.7 and was then adjusted to either pH 6.3 (for the stable suspension) or pH 7.8 (for the unstable suspension) using dilute NaOH solution.

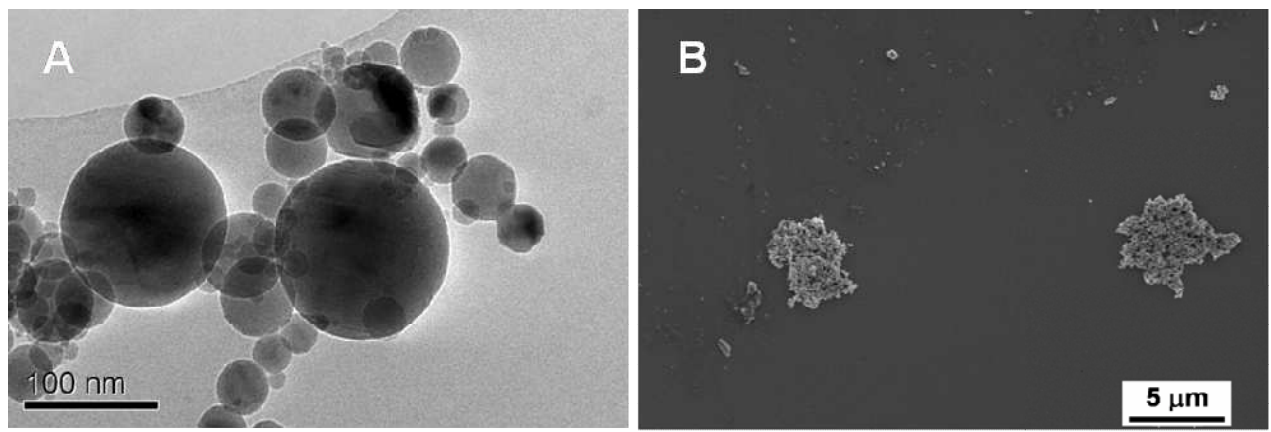

**Figure 1:** Images of Polydisperse Nanotek alumina nanopowder. The left image (A) is a TEM picture showing the individual nanoparticles within the as-received aggregated powder. The right image (B) is an SEM picture showing some of the aggregates found in the powder.

Size and zeta-potential measurements of the alumina-water suspensions were conducted using a Malvern Zetasizer-Nano instrument. For zeta potential measurements the samples were loaded to DTS1060 capillary cells. The iso-electric point (IEP) of the suspension was recorded as pH 8.9.

For size measurements the sample was loaded into the Zetasizer-Nano in a ZEN0112 disposable cuvette. Three consecutive measurements were taken and then averaged and plotted in figure 2. The first data point was obtained at t = 2 min, due to the typical setting-up time for the Zetasizer-Nano to begin a measurement. The data obtained on the stable suspension (pH 6.3) in general show that no change in the particle size took place, and the uncertainty in each measurement was small. This demonstrates that for a stable suspension, the Zetasizer-Nano may be used successfully at this particle concentration, even though the measured particle size in our case (~90 nm) is significantly larger than the value quoted by the manufacturers (46 nm), indicating the presence of aggregates. By contrast, the data obtained from the unstable suspension (pH 7.8) at the same particle concentration shows a decrease in particle size by more than a factor of 2 and the uncertainties for each measurement are very large. The apparent decrease in particle size is due to the settling of the larger particles and aggregates out of the field-of-view of the instrument. This demonstrates that Zetasizer-Nano measurements are erroneous even at 0.5 wt% for a highly polydisperse and unstable colloid.

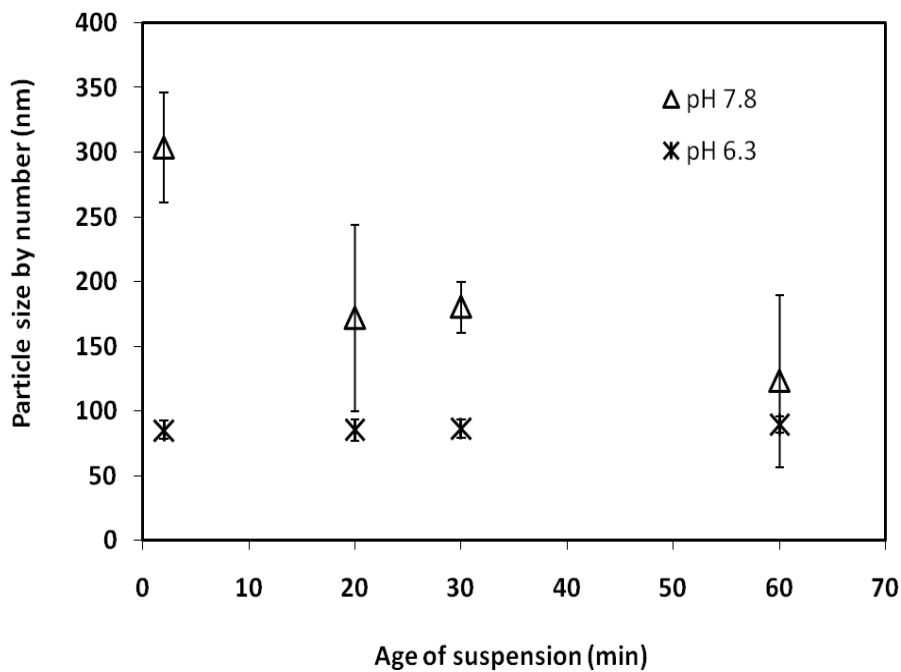

**Figure 2:** Zetasizer-Nano measurements for 0.5 wt% alumina-water suspensions. The crosses (pH 6.3) represent data from the stable suspension. The triangles represent data obtained from the unstable (pH 7.8) suspension.

**(b) Photographic study**

The results of gravity settling experiments carried out on the alumina-water suspensions were studied using digital photography. A sample from a large volume of each stirred suspension was extracted and placed into a 2.5 cm diameter transparent glass vial which was then carefully placed on a table in front of a black piece of cardboard for better photographic contrast. The volume of the sample was approximately 20 ml. A Nikon Coolpix8800 digital camera was used to photograph the vial at regular intervals, showing the movement of the solid-liquid interface with time. Two sets of photographs were produced, one for the stable (pH 6.3) suspension, the other for the unstable (pH 7.8) suspension. The time taken to transfer and to photograph the sample was less than one minute, thus making it possible to capture the early stages of the settling process.

**(c) Microscopy study on sessile droplets**

Microscopy studies were carried out on both evaporating droplets and fully dried deposits. The aim was to investigate the changes in aggregates during the evaporation process as well as the uniformity of the aggregates across the fully dried deposit.

A Nikon Eclipse TE2000 optical microscope (OM) was used to produce images of the evaporating droplets and the dry deposits. A few hundred microlitres of suspension was carefully extracted from the centre of the bulk solution using a Fisherbrand pipettor, and a 50 µl droplet was deposited onto glass slides that were cleaned with Decon and acetone to leave a contact angle of 20 – 25$^o$. The droplets were then allowed to dry under ambient conditions of 20 $^o$C and 40 % RH.

### (d) SAXS study

Unstable (pH 7.8) 0.5 wt% alumina-water suspensions were prepared for SAXS experiments at I22 beamline of the Diamond Light Source at Harwell, Oxfordshire, UK. More detailed descriptions on how SAXS may be used to measure particle size are to be found elsewhere (Glatter and Kratky 1982). There are two main advantages with the SAXS: firstly the I22 beamline has a detector set up at 6 m from the sample cell, meaning that very small angles (< 1°) may be investigated, and secondly, the intensity of the 1 Å X-rays is such that only 10 s are required to obtain a statistically reasonable count for a measurement.

A small sample from the freshly prepared and stirred suspension was transferred into the experimental SAXS cell. This cell was disk-shaped and 10 mm in diameter by 1 mm thickness, and required only 80 µl of liquid to fill it. The I22 beamline had a rectangular beam slit of 300 µm by 340 µm (height and width) centered on the experimental SAXS cell.

After data collection, a background subtraction was carried out and the data was put into one file using the in-house Dream software and SAXS Utilities software at Diamond.

## Results and discussion

### 1. Bulk Settling

*1.1 Photographic study*

Figure 3 shows both the stable (pH 6.3) and the unstable (pH 7.8) suspensions settling under gravity with time over a 30 minute period.

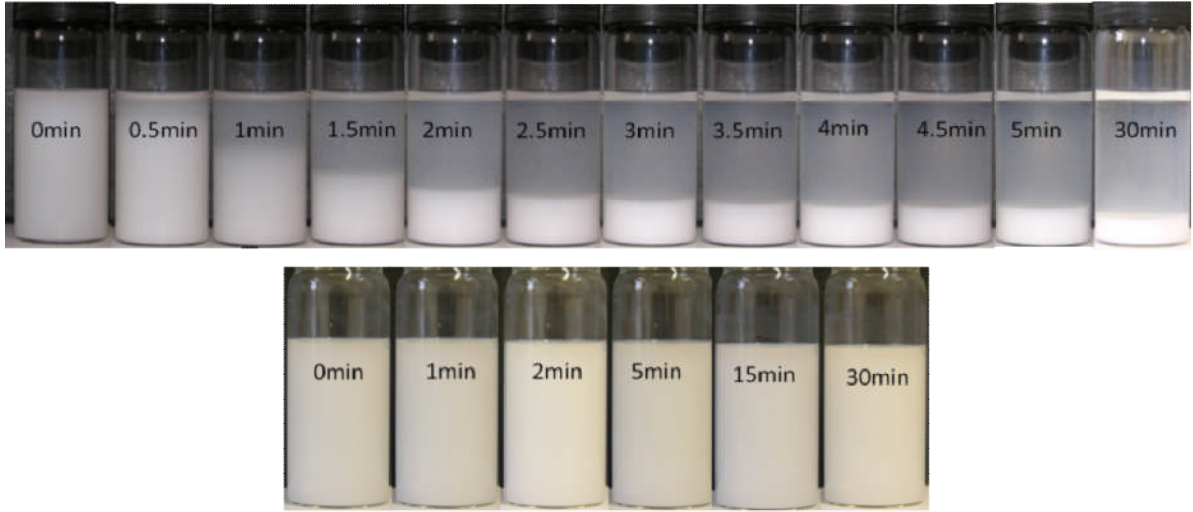

**Figure 3**: Gravity settling experiments of two aqueous 0.5 wt% alumina suspensions at pH 7.8 (top row) and pH 6.3 (bottom row).

The unstable suspension at pH 7.8 demonstrates rapid settling, particularly during the first two minutes of the experiment after which the change in bed height slows down considerably. The liquid above the particle bed becomes more transparent after 2 minutes suggesting most of the larger particles have settled out. But the liquid phase is not completely clear indicating that a few small particles must remain in it for a considerable period after this time. Meanwhile the stable suspension at pH 6.3 does not show any visible separation over the 30 minute period studied.

To examine the settling data more closely, the bed height ("interface height") versus the time of the experiment is plotted in Figure 4. The stable suspension bed height is always above the unstable suspension bed height since the entire 20 ml of solution always appeared filled, whereas the unstable suspension had a visible concentration gradient near the top of the solution at the measurement times. This suggests that there are some very large (i.e. rapidly settling) objects present in the unstable suspension.

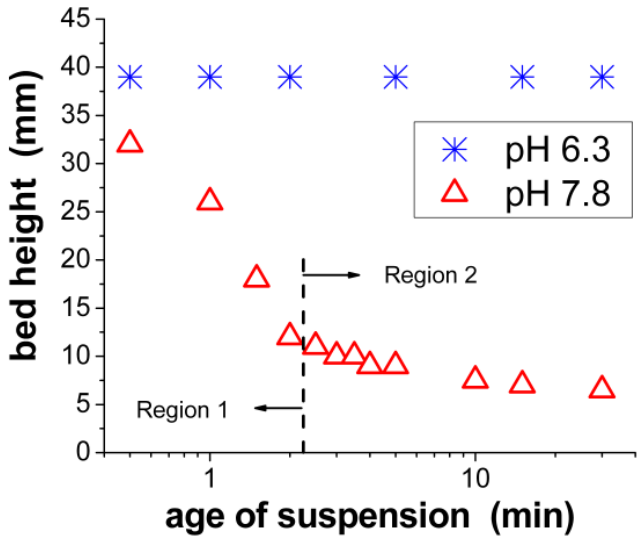

**Figure 4:** Measured bed heights of gravity settled 0.5 wt% aqueous alumina suspensions at the pH indicated.

Figure 4 can be split into two regions based on the rapidity of settling. The first region up to 2 minutes shows an initial onset period followed by rapid settling at approximately 46 mm/min. Region 2 demonstrates much slower settling at 4 mm/min. For the case of unhindered particles settling under gravity, Stokes law predicts the settling rate (U) for non-interacting spheres of diameter (d), particle density ($\rho_s$), liquid density ($\rho$) and liquid viscosity ($\mu$); $d = \sqrt{18\mu U /(\rho_S - \rho)g}$ . From this equation the mean particle size settling in Region 1 of Figure 4 is calculate as approximately 22 μm. In Region 2 the mean particle size is approximately 6 μm. Alumina particle density was taken as $\rho_s$=3970 kg/m$^3$. Clearly these calculated sizes are several times larger than any of the individual primary particles in the suspension. The existence of two distinct settling rates suggests that more than one process for settling is taking place within the suspension, which has also been previously reported by Turian et al(Turian et al. 1997). Apart from this, no further information on these processes can be extracted from these data alone.

*1.2 SAXS study*

Figure 5 presents the SAXS data for the unstable (pH 7.8) alumina suspension over a 15 minute period.

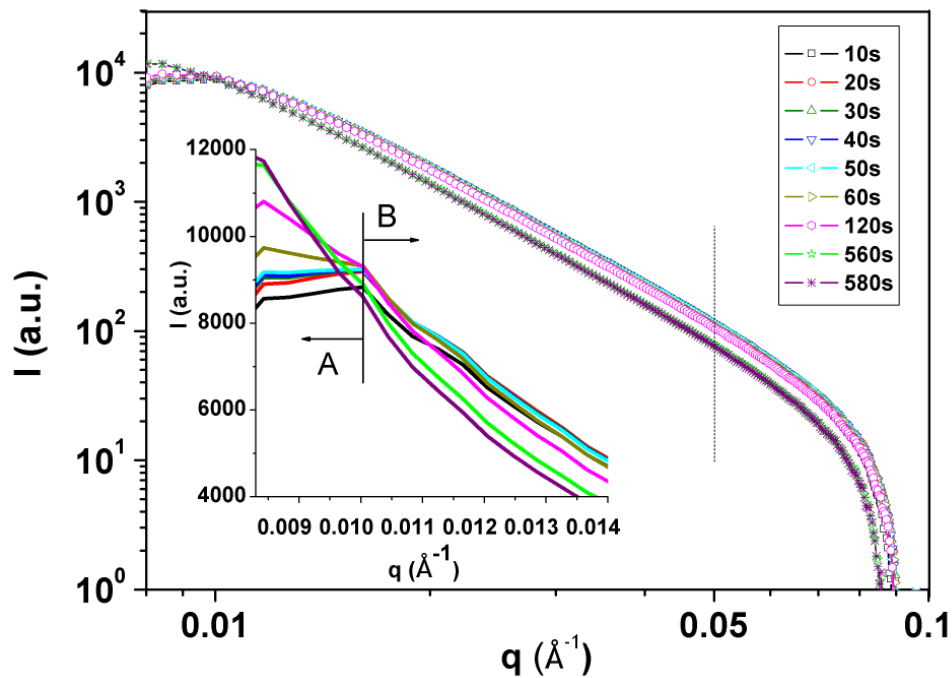

**Figure 5:** SAXS data for a 0.5 wt% alumina-water suspension at pH 7.8 at various times from the beginning of the experiment. The inset shows the rapid change in I at very small values of q (<0.014 Å$^{-1}$). A dotted line is drawn at q = 0.05 Å$^{-1}$ indicating where data analysis at larger q values was carried out.

The SAXS data presented in Figure 5 show two shoulders, one at q = 0.01 Å$^{-1}$, (see inset) the other at q = 0.08 Å$^{-1}$. The first shoulder at low q, which is seen to disappear with time, is likely to be a concentration effect due to the proximity of neighboring particles within the solution. Around this value of q (0.01 Å$^{-1}$) the SAXS measurement is sensing objects that are approximately 60 nm across. It may be seen from Figure 1 that this is near the median size of the alumina particles in the suspension. During the experiment, the smaller particles present in suspension will aggregate into clusters, many of which are likely to be in this size range, whereas most of the single particles of 60 nm will grow into much larger aggregates and settle out. This effect is observed as a smoothing out of the shoulder at q = 0.01 Å$^{-1}$ and an increase in the intensity at even smaller q. The ability to measure this change by SAXS is only possible because of the very high number of X-rays per cross-section that allows rapid sampling.

The other shoulder at q = 0.08 Å$^{-1}$ is observed to decrease with time, but remains prominent until the end of the experiment. This shoulder corresponds to objects that are approximately 8 nm across, which is the size of the smallest particles present in the suspension. Thus, when these particles coalesce into larger aggregates, no smaller particles can replace them so the measured intensity falls with time. The fact that this shoulder remains pronounced at the end of the experiments indicates that most of these smaller particles remain as single particles after the larger aggregates have settled out. This effect was visible in the settling experiment (Figure 3) where the solution above the settled bed is cloudy even at 30 minutes, suggesting the presence of many small particles that are more stable in solution.

In figure 6 we have taken SAXS intensity values at q = 0.008 Å$^{-1}$ (corresponding to the position of the maximum intensity measured at 580 s in Figure 5) and then plotted the relative change in these intensity values during the experiment.

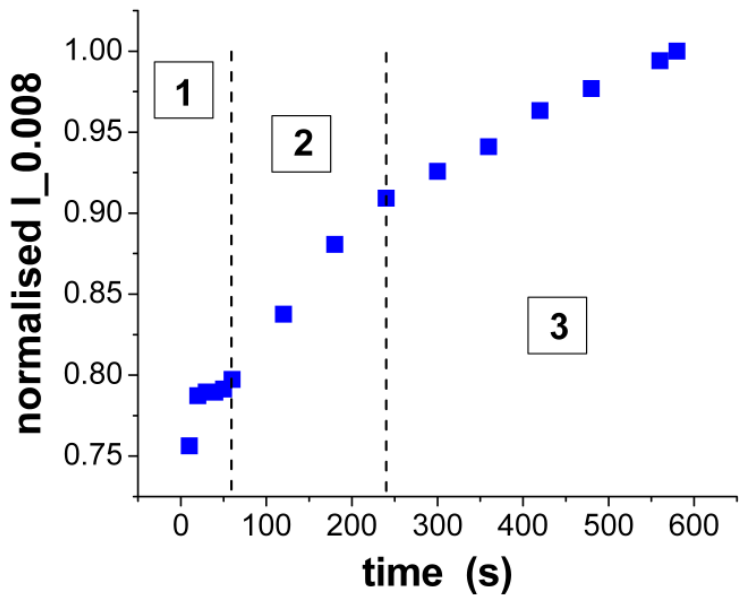

**Figure 6:** SAXS data showing the relative change in $I_{0.008}$ with time. The numbers refer to Region 1, 2 and 3 respectively (see text).

The SAXS data presented in Figure 6 show a general increase in $I_{0.008}$ with time throughout the experiment. In Region 1 there appears to be a rapid onset over the first 20 s before a slower increase takes place upto 60 s or so. It can be assumed that the very rapid removal of the largest aggregates is taking place during the first 20 s. Once these initial aggregates fell beyond the measurement zone, it seems that the creation of further objects in the size range detected by q = 0.008 Å$^{-1}$ (corresponding to ~ 80 nm) is much slower.

Between 60 s and 240 s (Region 2) a significant increase in $I_{0.008}$ takes place. This approximately matches the times found for the settling rate changes observed in the bed height measurements shown in Figure 4. Whilst it is too early to claim a direct link between bed height measurements and SAXS data, it is reasonable to deduce that there exists some relationship between the rate of creation of aggregates of a particular size (detected by SAXS) and the increased settling rate observed during the bed height measurements. Certainly there is a steady increase in the number of objects detected by $I_{0.008}$ during this period, before the onset of a slower increase in Region 3. The change from Region 1 to Region 2 can be described as follows; initially the 80 nm particles aggregated rapidly into micron-sized objects and settled out, allowing a bigger space for the smaller particles to come together and form more 80 nm particles. After only a short period, most remaining particles are much smaller than 80 nm and thus require a longer time to form an 80 nm aggregate. This marks the transition from Region 2 to Region 3. The fact that the relative intensity of X-rays continues to increase throughout the experimental period (10 minutes) suggests that many more 80 nm objects exist per unit volume at 10 minutes than initially.

Figure 7 shows a steady decrease in $I_{0.05}$ with time, suggesting the number of objects in this size range (~ 12 nm) reduces slowly with approximately two-thirds of the objects remaining unchanged after this period. This agrees with the settling data pictured in Figure 3 where the liquid above the settled bed is still cloudy after 10 minutes indicating that many small particles have not yet settled.

Note that at 600 s, $I_{0.05}$ is still decreasing steadily with no sign of levelling off. Monitoring the suspension for a longer period may give SAXS the capability of detecting the settling time required for much smaller particles than is currently possible with any other technique.

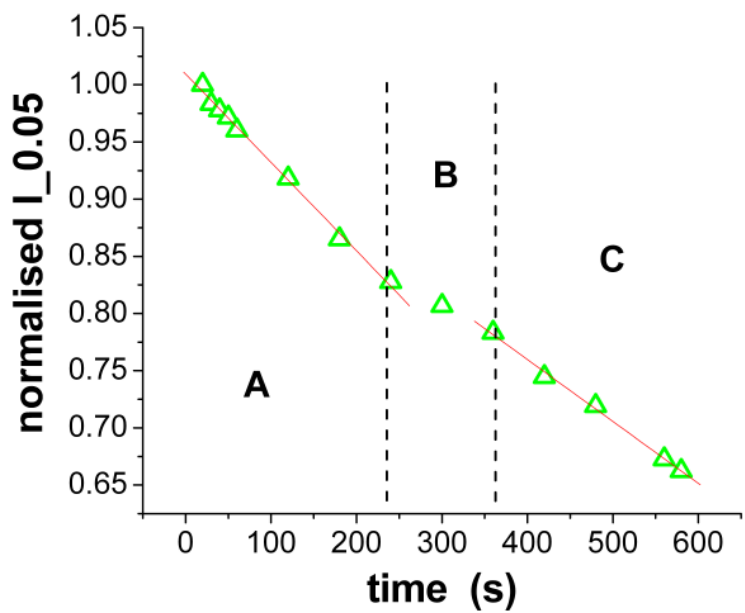

**Figure 7:** SAXS data showing the relative change in the intensity at q = 0.05 Å$^{-1}$. The straight lines are to guide the eye. Regions A, B and C are discussed in the text.

Unlike particles that are ~ 80 nm (Figure 6), no smaller particles are available to replace the 12 nm particles that have aggregated (Figure 7), and therefore the measured intensity of I$_{0.05}$ decreases with time. Figure 7 can be roughly split into 3 regions (A, B and C) of different gradients. This may suggest that more than one process is taking place, possibly linked to the availability of larger particles for the smaller particles to attach onto.

The information shown in Figures 5, 6 and 7 demonstrates the uniqueness of SAXS to examine unstable suspensions. In addition to the information obtained from the standard settling data (Figures 3 and 4), SAXS offers the capability to monitor the dynamics of the nanoparticles of different sizes while they are still in suspension. This is currently the only technique that allows the experimenter to access this kind of information for nanometre-sized particles in-situ. For comparison, the optical microscopy studies are presented below.

## 2. 2-D visualization of the unstable alumina suspension

*2.1 Study of evaporating droplets*

A 50μl droplet extracted from the unstable suspension at pH 7.8 was deposited onto a glass slide and examined by an optical microscope at room conditions. Images were taken through the microscope at regular intervals and are presented in Figure 8.

Initially at 0 s a limited number of dark objects are visible, with several of the objects being quite large (~ 5 to 10 µm in diameter). The fact that these large objects are in focus means that they have settled almost immediately onto the glass slide either because they happened to be near the bottom of the droplet, or because their settling time is very short.

At 30 s a much larger number of objects are present on the glass slide. These are likely to be a combination of medium-sized particles that require this time to settle out (~ 100 nm in diameter) plus some smaller particles that have aggregated together to form larger particles during this interval. If the image at 30 s is examined closely, it becomes apparent that only a few more micron-size objects have appeared, but many smaller objects (light grey areas) are visible.

The SAXS data presented above suggest that at this time only a small degree of aggregation has occurred, so most of these smaller objects comprising the light grey areas in Figure 8 must have already been formed during the preparation of the suspension.

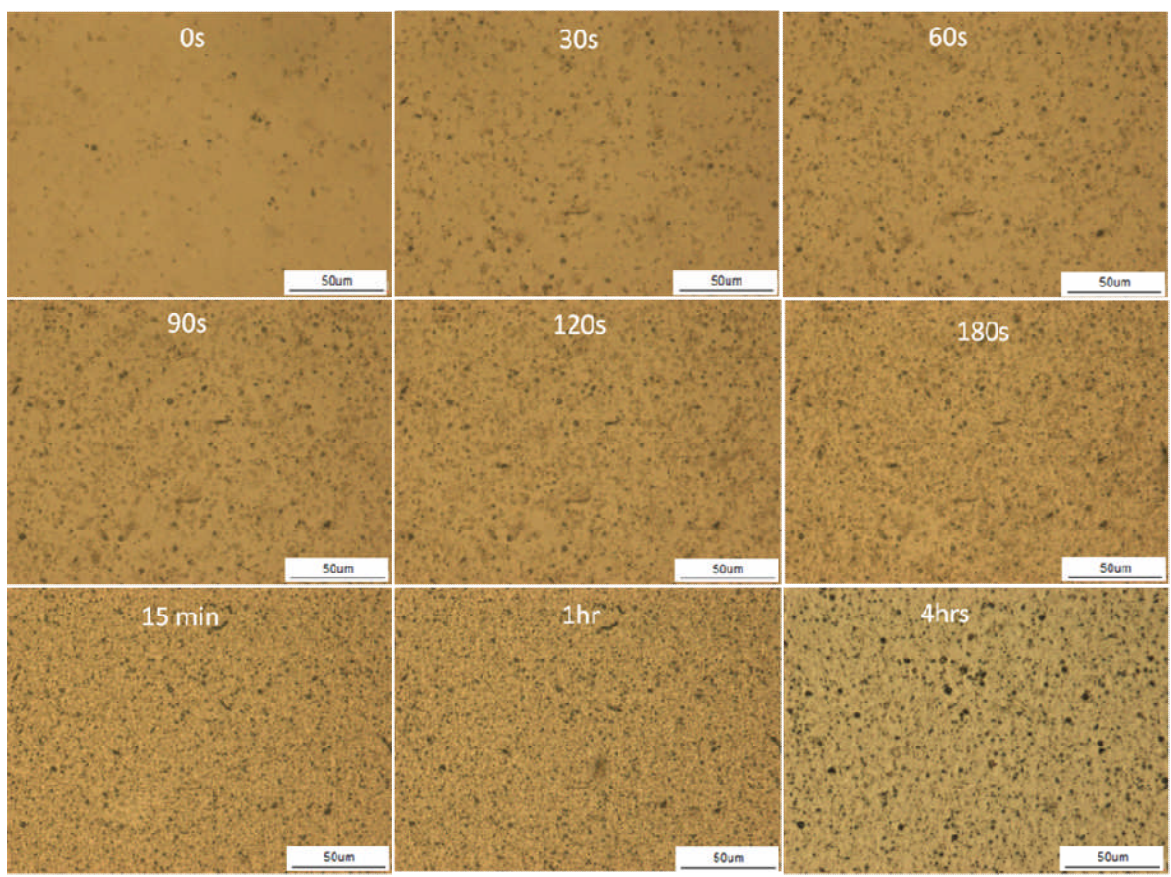

**Figure 8:** Optical microscopy images of a 50µl droplet taken from an unstable suspension of 0.5 wt% alumina at pH 7.8 as it evaporates under ambient conditions on a hydrophilic glass substrate. The times indicated on each image are measured from the moment the droplet was placed onto the glass slide.

Between 60 s and 180 s the percentage of the image covered by the light grey areas increases steadily, roughly agreeing with the period described above for the onset of aggregation and settling in the SAXS experiment on the same system (Figure 6), and with the rapid change in the bed height seen in the settling experiments (Figure 4).

Between 180 s and 15 minutes there are only small changes visible in the optical images. The number of large, black (micron-sized) objects has not changed noticeably, but it is possible that the percentage of the image covered by the light grey objects has increased slightly to form a more uniform coverage. This period corresponds approximately to Region 3 of the SAXS data in Figure 6 and to the slower changes in bed height observed in Figure 4. It is possible that the rate of aggregation of the smaller particles present in the suspension primarily determines the changes seen at these times, with only a certain fraction of the smaller particles reaching sufficient aggregate size to settle out.

Between 15 minutes and 1 hour no significant changes are observed, suggesting that the aggregation and settling of the medium-sized particles has all but stopped, and only the smallest (few nanometers) particles remain in solution.

The droplet was found completely dry after 4 hours, and only small differences are seen between the images at 4 hours and at 1 hour. Appreciate that the main difference is the image contrast has increased at 4 hours because the water is no longer present to defocus the light from the microscope. Most of the larger objects seen at 1 hour can be observed on the image at 4 hours with only a small degree of visible aggregation occurring during the final stages of drying as the liquid meniscus pulls some of the smaller particles on top of the larger particles. The patterns formed by the light grey areas at 1 hour are largely unchanged at 4 hours. This sequence of images taken during evaporation of the unstable aqueous alumina droplet shows the number of objects initially present in the droplet as well as what happens to each different size of object. In effect, the final dry deposit shows the result of the entire sedimentation and settling process in a two-dimensional domain, allowing examining the uniformity of the process across the bed.

*2.2 Study of dried deposits*

Droplets were extracted from the unstable suspension (pH 7.8) at different times during the settling process. The extractions were done near the centre of the suspension each time to simulate the measurement zone examined by the SAXS X-ray beam. Aim was to observe the structures formed across the deposit after drying.

The times indicted on the images in Figure 9 represent the moment the droplet was extracted from the bulk suspension. Thus particles that had already settled in the bottom were not extracted. Hence the particle concentration in each droplet illustrated in Figure 9 is not constant, but instead will decrease with time.

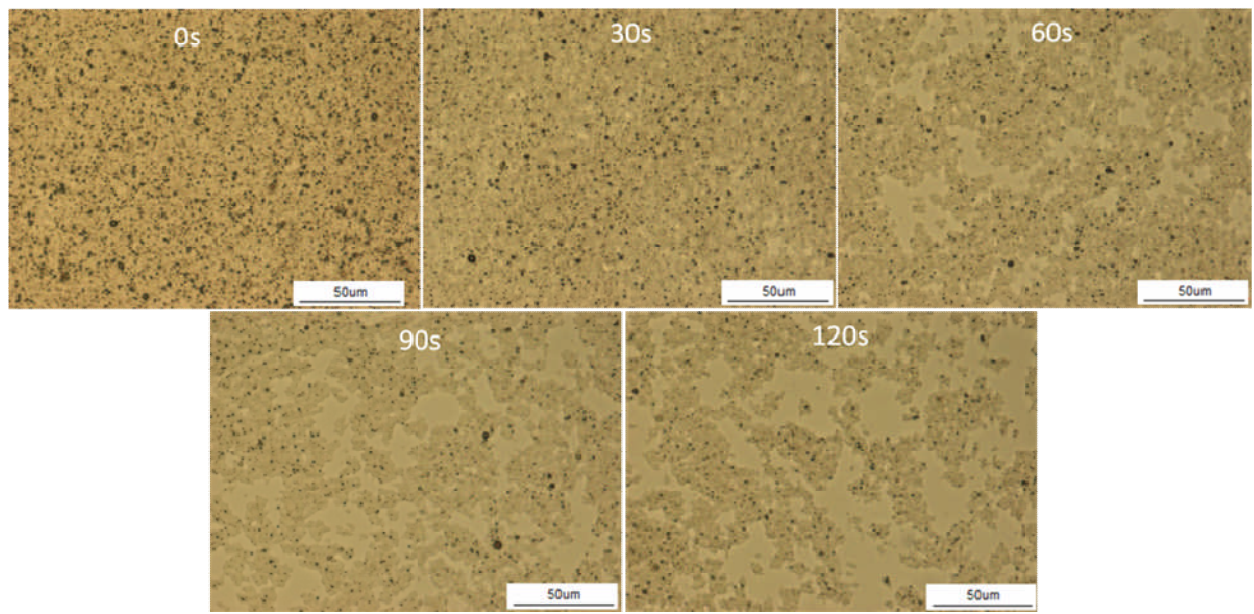

**Figure 9**: Optical microscopy images of dried deposits of an unstable alumina suspension at pH 7.8. The times labeled on each image correspond to the age of the suspension at which the sample was extracted to form the droplet.

The dry deposit formed from an evaporated droplet extracted as soon as the suspension was formed (0 s in Figure 9) shows a high concentration of particles of a wide variety of sizes, including plenty of large micron-size objects. At this stage, very few particles have had time to leave the centre of the suspension, so the total number of particles observed should represent the initial concentration (0.5 wt%).

The sample extracted at 30 s shows slightly fewer micron-size objects than the 0 s deposit, but now there are large sections of the image covered by light grey areas probably representing a bed of smaller particles. There are a few pin holes in this otherwise uniform grey layer. Recall the SAXS data in Figure 6 where only a small degree of aggregation had taken place within the suspension at this time. It is therefore reasonable to expect most particles to remain in the deposit after drying.

The 60 s image shows a sharp difference from the earlier images. There now exists significant holes in the particle layer formed, and there appear to be several quite large (tens of microns) aggregated zones across the image. This would suggest that a significant amount of settling has occurred within the suspension leaving a much lower particle concentration in the droplet at this time. Interestingly this corresponds to the onset of the steeper gradient seen in the SAXS data in Figure 6, but it occurs too soon to be observed by the standard settling test (Figure 4). The fact that the deposit forms large patches rather than a uniform grey colour suggests that most of the smaller particles present are attached to each other. Since only a little lateral movement occurred during the drying of these droplets (Figure 8), it is tempting to suggest that the large spaces present in the 60s image indicate the formation of these aggregates in solution rather than during the final stages of drying.

The images corresponding to 90, 120 and 180 s are quite similar. They show an increased loss of particles from the sampling zone due to the settling out of aggregates within the suspension resulting in a more fragmented deposit. However, the number of micron-size particles appears to be similar to that on 60s image. These are likely to represent aggregates formed by the smaller particles in the suspension, which was earlier showed by SAXS (Figure 7), occurred more slowly and continuously during the ageing of the suspension.

## Conclusions

Settling experiments were conducted with spherical, polydisperse, near-IEP, and dilute (0.5 wt%) aqueous alumina suspensions at room temperature. The aggregation and settling behaviour was examined using photography, optical microscopy and SAXS. From photographic studies and the corresponding settling curve, two settling regimes were identified; one showing the settling of very large (~ 22 μm) objects, the second showing a slower settling rate of relatively smaller (~ 6 μm) objects.

SAXS experiments revealed far more details of particle settling in the sample. The data was split into 3 regions that approximately matched (1) the time for the onset of settling, (2) the aggregation of smaller particles into micron-sized objects and (3) the slower growth of aggregates formed from the smallest particles. By analyzing the SAXS data more closely at $q = 0.05$ Å$^{-1}$, two distinct regions were identified that were ascribed to the high or low concentration of the larger particles. Use of the SAXS technique has thus provided a wealth of information for this highly unstable suspension that is otherwise not possible to obtain from any existing mechanism.

The study of the dry deposits by optical microscopy confirmed both the settling bed data and the non-uniformity in the bed changes as the sample ages. Clearly there were several very large objects initially present within the alumina suspension. The dry deposit examination made it possible to make an estimate of their population.

Moreover a significant number of micron-sized objects existed in the deposits taken at later times from the suspension.

In conclusion this paper has, for the first time, integrated gravitational settling bed tests with SAXS experiments and optical microscopy for a rapidly settling polydisperse alumina suspension. It was shown that each of these techniques complement one another in providing a complete picture of the activities occurring in a very complex colloid. In the future it would be interesting to study the behaviour of a bi-modal system whose settling rate was carefully controlled, and monitor the changes seen by SAXS to try and calibrate the intensity data. The influence of particle shape is another parameter worthwhile to investigate. These experiments would be very useful in predicting the complex settling and aggregation process of industrial particles at high concentrations. This paper represents the first step in this direction.

Acknowledgements

We acknowledge the help provided by Prof Nick Terrill, Dr Marc Malfois, Dr Claire Pizzey, and Dr Jen Hiller of the I22 beamline in the Diamond Light Source, and Dr Richard Heenan of ISIS, UK. We are grateful to the EPSRC for grants: EP/F023014/1 and EP/F000464/1. We also acknowledge Dr Jian Jin for his helping hand in SAXS experiments, and Dr Mohammed Rehan for TEM images.